\documentclass[a4paper,11pt]{article}
\usepackage{jheppub} 
\usepackage[utf8]{inputenc}
\usepackage{graphicx}

\usepackage{wrapfig}
\usepackage{caption}
\usepackage{subcaption}
\usepackage{setspace}
\usepackage{amsmath}
\usepackage{amsfonts, bm}
\usepackage{commath}
\usepackage{amsthm}
\usepackage{dsfont}
\usepackage{color}
\usepackage{braket}
\usepackage{color,soul}
\usepackage{appendix}
\usepackage{amssymb}
\allowdisplaybreaks
\usepackage[dvipsnames]{xcolor}

\usepackage{lineno}

\usepackage{mathtools}

\title{Post-Markovian master equation \`{a} la microscopic collisional model}

\author[a, b]{Tanmay Saha,}
\author[a, b]{Sahil,}
\author[a, b]{K. P. Athulya,}
\author[a, b]{Sibasish Ghosh}
\affiliation[a]{Optics \& Quantum Information Group, The Institute of Mathematical Sciences, CIT Campus, Taramani, Chennai 600113, India}
\affiliation[b]{Homi Bhabha National Institute, Training School Complex, Anushakti Nagar, Mumbai 400085, India}
\emailAdd{tanmaysaha32@gmail.com}
\emailAdd{sahil402b2@gmail.com}
\emailAdd{athulyakp@imsc.res.in}
\emailAdd{sibasish@imsc.res.in}

\abstract{We derive a completely positive post-Markovian master equation (PMME) from a microscopic Markovian collisional model framework, incorporating bath memory effects via a probabilistic single-shot measurement approach. This phenomenological master equation is both analytically solvable and numerically tractable. Depending on the choice of the memory kernel function, the PMME can be reduced to the exact Nakajima-Zwanzig equation or the Markovian master equation, enabling a broad spectrum of dynamical behaviors. We also investigate thermalization using the derived equation, revealing that the post-Markovian dynamics accelerates the thermalization process, exceeding rates observed within the Markovian framework. Our approach solidifies the assertion that \textit{``collisional models can simulate any open quantum dynamics"}, underscoring the versatility of the models in realizing open quantum systems.
}

\begin{document}
\maketitle
\flushbottom

\section{Introduction}
\label{intro}
The fundamental interest of the theory of open quantum systems is to investigate a quantum system in contact with an external environment or bath. Studying open quantum systems \cite{breuer02, Rivas2012Huelga, alicki2007quantum} enables precise predictions regarding the dynamical behavior of a quantum system interacting with its environment. There are primarily two distinct approaches to addressing such systems: the quantum Langevin equation (QLE) \cite{10.1063/1.1704304, PhysRevLett.46.1, PhysRevA.37.4419} method, which deploys effective Heisenberg equations to capture the reduced dynamics of the system of interest, and the density operator formalism, where the system's dynamics is governed by an appropriate master equation (ME) \cite{breuer02, Rivas2012Huelga}. Notwithstanding the extensive use of the master equation approach to describe various systems, including those in quantum optics, atomic physics, and condensed matter physics to name a few, deriving such a closed-form master equation from first principles remains challenging.  This difficulty arises from limited knowledge about the environment and its precise interaction with the system. By neglecting the memory effects of the environment --- employing the Markovian approximation via Born-Markov and rotating-wave approximations --- one typically obtains the celebrated Gorini-Kossakowski-Sudarshan-Lindblad (GKSL) ME \cite{KOSSAKOWSKI1972247, 10.1063/1.522979, Lindblad1976, breuer02, Rivas2012Huelga}, which guarantees unconditionally completely positive and trace preserving (CPTP) dynamics \cite{kraus1983states}. While the Markovian ME is both analytically solvable and numerically tractable, it remains primarily of theoretical interest, as real-world environments generally exhibit non-negligible memory effects. Along comes non-Markovian MEs, a wide range of realistic phenomena involving memory effects \cite{NM3, b-review, rivas14quantum}. In most cases, however, non-Markovian (NM) MEs \cite{nakazima, zwanzig} are neither analytically solvable nor numerically tractable. Additionally, complete positivity is generally not guaranteed throughout the evolution, rendering non-CPTP (\textit{i.e.,} unphysical) dynamics in certain regimes \cite{PhysRevA.81.042103, dariusz-CPT}.

In recent years, the \textit{post-Markovian master equation} approach has become increasingly significant in the theory of open quantum systems \cite{breuer02, Rivas2012Huelga}, as it aims to simultaneously address $(i)$ the generalization of the GKSL ME to incorporate environmental memory effects, $(ii)$ the preservation of complete positivity, and $(iii)$ the need to maintain both analytical and numerical solvability. In the literature, a handful of \textit{post-Markovian master equations} (PMMEs) have been put forth \cite{Shibata1977Takahashi, Chaturvedi1979, PhysRevA.50.3650, PhysRevLett.77.3272, PhysRevA.55.4636, PhysRevA.59.1633, PhysRevLett.82.1801, YU2000331, PhysRevA.64.033808, PhysRevA.66.012108, PhysRevA.70.010304, PhysRevA.70.012106}; nevertheless, almost none of them satisfy all three of the aforementioned attributes at the same time. In \cite{Shabani-Lidar2005PMME}, Shabani and Lidar (SL) introduced a new PMME, formulated via a measurement approach, which fulfills the previously outlined objectives. They showed \cite{Shabani-Lidar2005PMME} that this PMME interpolates between the generalized measurement interpretation of the exact Kraus operator sum map \cite{kraus1983states} and the continuous measurement interpretation of the Markovian dynamics \cite{PhysRevLett.68.580, N_Gisin_1992, N_Gisin_1993, RevModPhys.70.101}.

Over the past two decades, collisional models (CMs) have acquired substantial importance in advancing research on open quantum systems \cite{ziman1, ziman2, CICCARELLO20221, collisional-review2}. CMs serve as effective toy models for addressing conceptual hurdles in the theory of open quantum systems that might be difficult to tackle using standard system-bath microscopic models. In the standard CM framework, a system undergoes sequential interactions with a series of identically prepared ancillas that collectively constitute the environment \cite{rau}. Since CMs dynamics involve interactions with only a few degrees of freedom at a time, they are simpler and easier to control than conventional approaches to open quantum systems. When the ancillas are initially uncorrelated and each collides with the system only once, CM inherently induces Markovian dynamics for the system. In the continuous time limit, this framework yields dynamics governed by GKSL ME, achieved without the need for additional approximations \cite{brun, ziman3, ziman4}. The simplicity and inherently discrete nature of CMs make them advantageous for investigating non-Markovian dynamics when the basic model is modified to incorporate a memory mechanism. Non-Markovian dynamics may emerge through various modifications, such as incorporating ancilla–ancilla interaction \cite{Saha_2024, PhysRevA.87.040103, Ciccarello_2013, paternostro-strategy, giovanetti-all, lorenzo-intra, strunz-intra, cakmak-intra}, utilizing initially correlated bath \cite{ziman5, bernardes-1, bernardes-2, vega-1, filippov-1}, employing a composite collisional model \cite{palma-composite}, or allowing the system to undergo multiple collisions with each ancilla \cite{grimsmo, grimsmo-2}. In the setting of NM CM, an ME was posed by Ciccarello, Palma, and Giovannetti \cite{PhysRevA.87.040103} by introducing ancilla–ancilla interaction, again without requiring any additional assumptions.

Owing to the remarkable success of CMs in deriving MEs for both Markovian and various non-Markovian dynamics \cite{CICCARELLO20221, PhysRevA.87.040103}, a prevailing view in the literature suggests that ``\textit{collisional models can simulate any open quantum dynamics}''. At this juncture, a pertinent question arises: can a post-Markovian master equation be derived with the aid of collisional models? In this paper, we answer this question affirmatively. In particular, we derive a completely positive post-Markovian master equation from a microscopic Markovian collisional model framework, incorporating bath memory effects via a probabilistic single-shot measurement approach. We then obtain the analytical solution for this phenomenological ME. In the qubit case, where both the system and ancillas are qubits, we numerically compute the solution. We establish a necessary and sufficient condition for complete positivity based on the memory kernel function. Thus, the derived PMME simultaneously satisfies all three of the previously stated qualities. By selecting appropriate memory kernel function, the PMME can be mapped to either the exact Nakajima-Zwanzig equation \cite{nakazima, zwanzig} or the Markovian ME. Our derivation of the PMME through CM takes us one step further towards establishing the coveted goal of proving the versatility of the models in realizing any physical open system dynamics. Utilizing the derived ME, the process of thermalization --- a central focus of quantum thermodynamics \cite{vinjanampathy16quantum, binder18book} --- is subsequently investigated. In the qubit case, it is shown that confining the maximum weight of the memory kernel function to the early stages of the dynamics leads to a faster thermalization compared to the Markovian scenario.

The rest of the paper is organized as follows: In Section \ref{preli}, we briefly review the basic framework of the collisional model. In Section \ref{derivation}, we derive the post-Markovian master equation using a collisional model. Section \ref{solSection} provides its solution and examines the condition ensuring the complete positivity of the dynamics. Subsequently, in Section \ref{thermalSection}, we study qubit thermalization within the framework of post-Markovian dynamics, providing a qualitative comparison with the thermalization process under Markovian dynamics. Finally, Section \ref{conclusion} presents the conclusions.
\section{Collisional model}
\label{preli}
\begin{figure*}[ht!]
	\centering
	\begin{minipage}{0.45\textwidth}
		\centering
            \hspace{-0.8cm}
		\includegraphics[width=7.5cm, height=8cm]{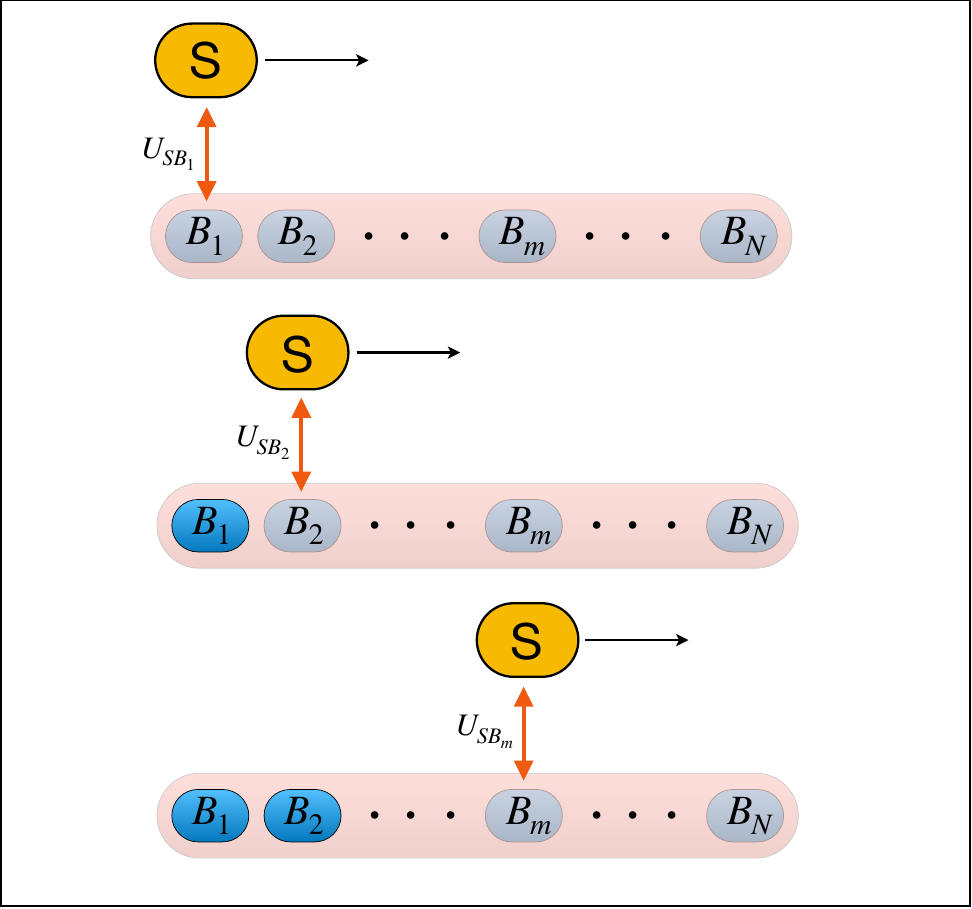}
		\subcaption{}
		\label{markovian}
	\end{minipage}
	\begin{minipage}{0.45\textwidth}
		\centering
		\includegraphics[width=7.5cm, height=8cm]{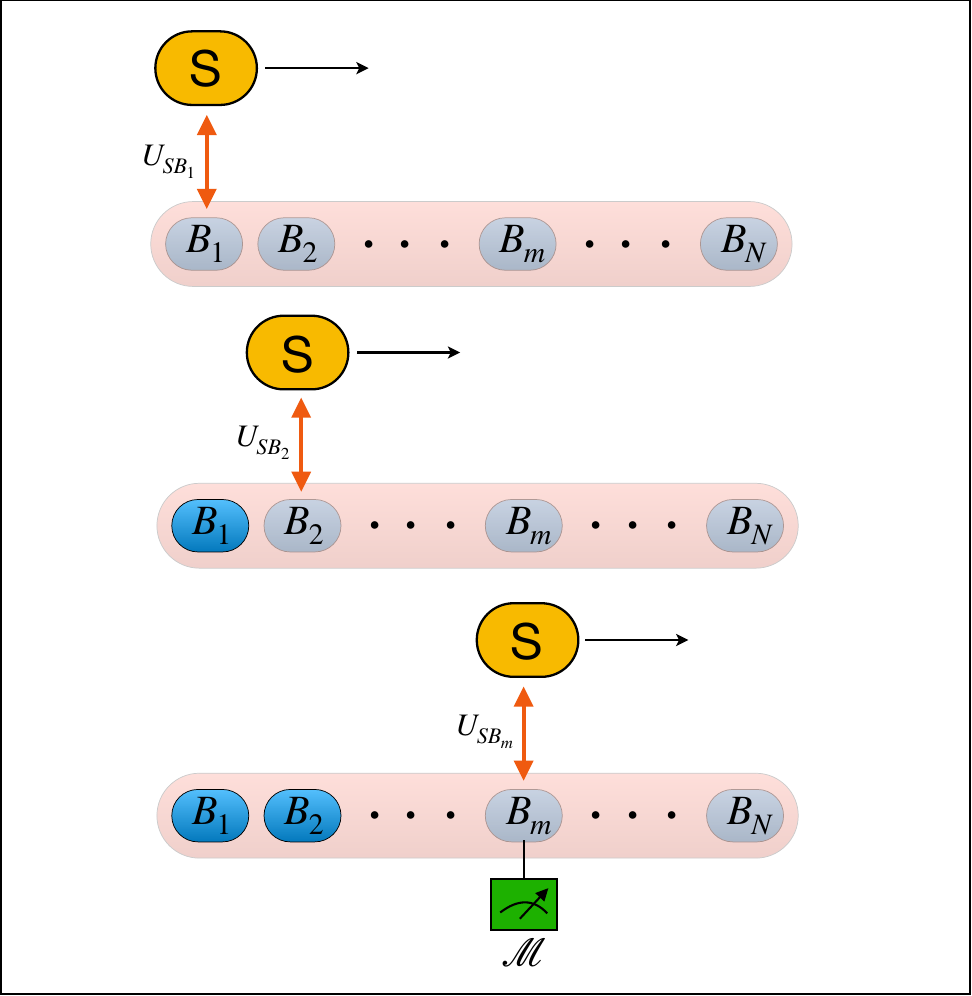}
		\subcaption{}
		\label{post-markovian}
	\end{minipage}
	\caption{ Schematic of (a) Markovian collisional model (b) Markovian collisional model modified with measurement on the $m$th ancilla.}
	\label{Schematic}
\end{figure*}

Collisional model is instrumental for the derivation of our master equation; therefore, we begin this section with a brief overview of its fundamental concepts. In CMs, the open quantum system 
under study interacts with an environment or bath, modeled as a collection of smaller and identical sub-systems, typically referred to as ancillas. There are no constraints on the dimensionality of the Hilbert space for either the system or the ancillas. The interaction between the system and the environment is then described by the repeated, sequential interactions (collisions) of the system with each ancilla.  In the simplest case, it is assumed that no initial correlations exist between the system and the individual units of the environment. Figure \ref{markovian} illustrates the simplest collisional model, in which the system $(S)$ interacts sequentially with a series of non-interacting ancillas, denoted as $B_{1}$, $B_{2},\cdots,$ representing the environment. 
Each ancilla collides with the system only once. Assuming the initial state of the system is $\rho_{S}(0)$ and each ancilla is prepared in the state $\eta$, the initial state of the composite system, consisting of the system and the bath, is expressed as 
\begin{equation}
   \sigma_{SB}(0) = \rho_{S}(0) \otimes\eta \otimes \eta \otimes \cdots
\end{equation}
The collision between the system $S$ and the $m^{\text{th}}$ ancilla $B_{m}$ is governed by the unitary operator $U_{SB_{m}}$. Assuming each collision lasts for the same duration $\tau$, the unitary operator for the $m^{\text{th}}$ collision is described by 
\begin{equation}
    U_{SB_{m}} \equiv U_{m}(\tau) = e^{-i\left(H_{S}+H_{B_{m}}+V_{m}\right)\tau},
\end{equation}
here, $H_{S}$ denotes the system Hamiltonian, $H_{B_m}$ is the Hamiltonian of the $m^{\text{th}}$ ancilla, and $V_{m}$ represents the interaction Hamiltonian between the system and the $m^{\text{th}}$ ancilla.

After $N$ collisions, the state of the composite system is given by
\begin{equation}
    \sigma_{SB}(N\tau) = U_{N}(\tau) \cdots U_{1}(\tau) \sigma_{SB}(0) U_{1}^{\dagger}(\tau) \cdots U_{N}^{\dagger}(\tau).
\end{equation}
The state of the open system is obtained by tracing out the environmental degrees of freedom, resulting in
\begin{align}
    \rho_{S}(N\tau) = {\rm Tr_{B}} \left\{\sigma_{SB}(N\tau)\right\} &= {\rm Tr_{B_{N}}} \left\{ U_{N}(\tau)\left(\rho_{S}((N-1)\tau)\otimes\eta\right)U_{N}^{\dagger}(\tau)\right\}\nonumber\\
    &\equiv \xi(\tau) \left[\rho_{S}((N-1)\tau)\right],
\end{align}
where $\xi(\tau)\left[\rho\right] = {\rm Tr_{B_{N}}} \left\{ U_{N}(\tau) \left(\rho \otimes \eta\right) U_{N}^{\dagger}(\tau) \right\}$ represents a completely positive map (also known as the dynamical map) that describes the transformation of the system's state. Here the map $\xi(\tau)$ is determined by the unitary describing each collision,  as well as the initial ancilla state $\eta$. 
Since both the unitary and the initial ancilla state are assumed to be the same for each collision in this simplest collisional model scenario, the state of the system $S$ after $N$ collisions is described by the composition of $N$ identical maps $\xi(\tau)$, as follows:
\begin{equation}
    \rho_{S}(N\tau) = \left\{\xi(\tau)\right\}^{N}\left[\rho_{S}(0)\right] \equiv \Lambda(N\tau)\left[\rho_{S}(0)\right],
\end{equation}
where $\Lambda(N\tau) = \left\{\xi(\tau)\right\}^{N}$. This particular exponential dependence of $\Lambda(N\tau)$ on the collisional map $\xi(\tau)$ gives rise to a discrete version of the semigroup property for $\Lambda(N\tau)$ given by
\begin{equation}
    \Lambda(N\tau) = \Lambda\big((N-m)\tau\big) \Lambda(m\tau),
\end{equation}
for any integer $m$ such that $0\leq m \leq N$. For this reason, this simplest version of the collisional model is commonly referred to as the Markovian collisional model.
\section{Derivation of the post-Markovian master equation}
\label{derivation}
To derive the master equation, we resort to the Markovian CM framework, as depicted in figure~\ref{post-markovian}. In the standard (Markovian) CM paradigm, we introduce a probabilistic single-shot measurement on ancilla in between the process. Here, `probabilistic' indicates that the choice of the ancilla on which the measurement is performed is determined by a specified probability distribution. This fact will be reflected by the phenomenologically introduced memory kernel function (MKF) in the derived ME. Since we are performing a non-selective measurement, it plays an auxiliary role; however, the probabilistic nature of the measurement is crucial in inducing the bath memory effects in the dynamics of the system of interest. In the following, we derive the ME starting from the collisional model.

We consider an arbitrary initial state of the system $(S)$, denoted as $\rho_{S}(0)$, and an arbitrary initial state of the $j^{\text{th}}$ ancilla $(B_{j})$, represented by $\eta_{j}=\ket{\chi_{j}}\bra{\chi_{j}}$. Although all initial ancilla states are identical, \textit{i.e.,} $\eta_{j}=\eta=\ket{\chi}\bra{\chi}$ for all $j$, we retain the subscript for clarity. Assuming the bath consists of $N$ ancillas (with $N$ being large), the composite initial state of the system and bath is given by $\rho_{SB}(0)=\rho_{S}(0)\otimes\eta^{\otimes N}$. Now, consider a scenario in which the CM is evolved with an arbitrary unitary $U(\tau)$ up to the $m^{\text{th}}$ ancilla, where $\tau$ denotes the duration of the unitary interaction. A non-selective projective measurement is then performed on the $m^{\text{th}}$ ancilla, after which the remainder of the CM is evolved. Assume that $\{\ket{M_{m}^{l}}\}$ forms an orthonormal basis for the measurement on the $m^{\text{th}}$ ancilla, such that it satisfies the following conditions:
\begin{equation}
\sum_{l}\ket{M_{m}^{l}}\bra{M_{m}^{l}}=\mathds{1}_{d_{m}}~~~\text{and}~~~\braket{M_{m}^{l}|M_{m}^{l'}}=\delta_{ll'},\label{OrthoCondition}
\end{equation}
where $d_{m}$ denotes the dimension of the $m^{\text{th}}$ ancilla.

Accordingly, the composite state of the system and bath is given by
\begin{align}
    \rho_{SB}(N\tau) =& \sum_{l} U_{N}(\tau) \cdots \ket{M_{m}^{l}}\bra{M_{m}^{l}} \mathcal{U_{M}} U_{m}(\tau) \cdots U_{1}(\tau) \Big\{\rho_{S}(0) \otimes \eta_{1} \otimes \eta_{2} \otimes\cdots\nonumber\\
    &\cdots \otimes \ket{\chi_{m}}\bra{\chi_{m}} \otimes \cdots \otimes \eta_{N}\Big\} U_{1}^{\dagger}(\tau) \cdots U_{m}^{\dagger}(\tau) \mathcal{U_{M}^{\dagger}} \ket{M_{m}^{l}}\bra{M_{m}^{l}} \cdots U_{N}^{\dagger}(\tau) \nonumber\\
    =& \sum_{l} U_{N}(\tau) \cdots \big\{A_{m}^{l}(\tau)\big\} \cdots U_{1}(\tau) \Big\{\rho_{S}(0) \otimes \eta_{1} \otimes \eta_{2} \otimes\cdots\nonumber\\
    &\cdots \otimes \ket{M_{m}^{l}}\bra{M_{m}^{l}} \otimes \cdots \otimes \eta_{N}\Big\} U_{1}^{\dagger}(\tau) \cdots \big\{A_{m}^{l}(\tau)\big\}^{\dagger} \cdots U_{N}^{\dagger}(\tau),
\end{align}
where $\mathcal{U_{M}}$ is the unitary corresponding to the pre-measurement interaction between the system and the $m^{\text{th}}$ ancilla, and $A_{m}^{l}(\tau) = \bra{M_{m}^{l}} \mathcal{U_{M}} U_{m}(\tau) \ket{\chi_{m}}$ is the Kraus operator acting on the system, satisfying the completeness relation:
\begin{equation}
    \sum_{l} \big\{A_{m}^{l}(\tau)\big\}^{\dagger} \big\{A_{m}^{l}(\tau)\big\} = \mathds{1}_{d_{S}},\label{KrausCompleteness}
\end{equation}
where, $d_{S}$ is the dimension of the system. The state of the system is obtained by tracing out the degrees of freedom of the bath,
\begin{align}
    \rho_{S}(N\tau) &= {\rm Tr}_{B} \Big\{ \rho_{SB}(N\tau)\Big\}\nonumber\\
    &= \sum_{l}\Lambda \big((N-m)\tau \big) \big\{A_{m}^{l}(\tau)\big\} \rho_{S}\big((m-1)\tau\big)\big\{A_{m}^{l}(\tau)\big\}^{\dagger}\nonumber\\
    &= \Lambda \big((N-m)\tau \big) \Tilde{\rho}_{S}\big(m\tau\big),\label{DeterministicMeasurement}
\end{align}
where $\Lambda(\tau)$ is the dynamical map defined in the previous section, and $\Tilde{\rho}_{S}\big(m\tau\big)$ represents the post-measurement state, given by
\begin{align}
    \Tilde{\rho}_{S}\big(m\tau\big) &= \sum_{l} \big\{A_{m}^{l}(\tau)\big\} \rho_{S}\big((m-1)\tau\big)\big\{A_{m}^{l}(\tau)\big\}^{\dagger}\nonumber\\
    &= \mathcal{E}(\tau)\big\{\rho_{S}\big((m-1)\tau\big)\big\},\label{measurementCPTP}
\end{align}
here, $\mathcal{E}(\tau)$ is the CPTP map associated with the measurement.

Equation (\ref{DeterministicMeasurement}) describes the system's state at time $t=N\tau$ for a deterministic measurement on the $m^{\text{th}}$ ancilla. However, deriving the PMME calls for a probabilistic measurement approach. The probability of a bath measurement occurring at the $m^{\text{th}}$ ancilla depends on the intrinsic memory properties of the bath, which are captured by the phenomenologically introduced memory kernel function $k\big((N-m)\tau, N\tau\big)$. The final state of the system (at $t=N\tau$) is obtained by averaging over different weighted measurements on ancillas, described as
\begin{align}
    \rho_{S}(N\tau) &= \sum_{m=1}^{N}k\big((N-m)\tau, N\tau\big)\Lambda \big((N-m)\tau \big) \Tilde{\rho}_{S}\big(m\tau\big)\nonumber\\
    &= \sum_{m=1}^{N}k\big(m\tau, N\tau\big) \Lambda\big(m\tau\big) \Tilde{\rho}_{S}\big((N-m)\tau\big),\label{probabilisticMeasurement}
\end{align}
with MKF satisfying the normalization condition $\sum_{m=1}^{N}k\big(m\tau, N\tau\big)=1$.
To obtain the ME for the system of interest, the primary requirement is to transition from the discrete case to the continuous-time limit. For the passage to the continuum limit, we let the number of collisions approach infinity, \textit{i.e.,} $m,N\to\infty$, while allowing the collision time to approach zero, \textit{i.e.,} $\tau\to 0$, such that the times $t'=m\tau$ and $t=N\tau$ remain finite. In this limit, equation (\ref{measurementCPTP}) then takes the form
\begin{equation}
    \Tilde{\rho}_{S}(t') = \mathcal{E} \big\{\rho_{S}(t')\big\}.
\end{equation}
Starting from equation (\ref{probabilisticMeasurement}) and subtracting the corresponding expression for $N-1$, then taking the continuous-time limit, yields
\begin{equation}
    \dfrac{\partial\rho_{S}(t)}{\partial t} = \int_{0}^{t} dt' \Bigg[k(t',t)\Lambda(t')\dfrac{\partial\Tilde{\rho}_{S}(t-t')}{\partial (t-t')}+\dfrac{\partial k(t',t)}{\partial t}\Lambda(t')\Tilde{\rho}_{S}(t-t')\Bigg].\label{firstForm}
\end{equation}
Assuming the intermediate evolutions to be Markovian, we set $\Lambda(t)=e^{\mathcal{L}t}$, where $\mathcal{L}$ is a Lindbladian \cite{10.1063/1.522979, Lindblad1976}. With this assumption, equation (\ref{firstForm}) becomes
\begin{equation}
    \dfrac{\partial\rho_{S}(t)}{\partial t} = \int_{0}^{t} dt' \Bigg[k(t',t) e^{\mathcal{L}t'}\circ\mathcal{E}\circ\mathcal{L}+\dfrac{\partial k(t',t)}{\partial t} e^{\mathcal{L}t'}\circ\mathcal{E}\Bigg]\rho_{S}(t-t'),\label{secondForm}
\end{equation}
where ``$\circ$" represents the composition of superoperators. In deriving the last equation, we used $ \Tilde{\rho}_{S}(t-t') = \mathcal{E}\circ\Lambda(t-t')\big\{\rho_{S}(0)\big\} = \mathcal{E}\circ e^{\mathcal{L}(t-t')}\big\{\rho_{S}(0)\big\}$. The left-hand side of the above equation is traceless, as ${\rm Tr}\dfrac{\partial \rho_{S}(t)}{\partial t}=\dfrac{\partial}{\partial t}{\rm Tr}\rho_{S}(t)=0$. Given that both $e^{\mathcal{L}t'}$ and $\mathcal{E}$ are trace-preserving operators, the trace of the first term in equation (\ref{secondForm}) is
\begin{align}
    {\rm Tr}\bigg[\int_{0}^{t} dt' k(t',t) e^{\mathcal{L}t'}\circ\mathcal{E}\circ\mathcal{L}\rho_{S}(t-t')\bigg]&=\int_{0}^{t} dt' k(t',t){\rm Tr}\bigg[e^{\mathcal{L}t'}\circ\mathcal{E}\circ\mathcal{L}\rho_{S}(t-t')\bigg]\nonumber\\
    &=\int_{0}^{t} dt' k(t',t){\rm Tr}\bigg[\mathcal{L}\rho_{S}(t-t')\bigg]=0,\label{firstTermTrace}
\end{align}
since it can be shown that a Lindbladian $\mathcal{L}$ acting on any operator $X$ satisfies
\begin{align}
    {\rm Tr}[\mathcal{L}X]={\rm Tr}\Big[-\dfrac{i}{\hslash}[H,X]+\dfrac{1}{2}\sum_{\alpha}a_{\alpha}\Big([L_{\alpha},X L_{\alpha}^{\dagger}]+[L_{\alpha}X,L_{\alpha}^{\dagger}]\Big)\Big]=0,
\end{align}
where $H$ is the effective Hamiltonian of the system on which $\mathcal{L}$ acts, $L_{\alpha}$ are Lindblad operators, and $a_{\alpha}\geq 0$ describe the decoherence and/or dissipation rates. Thus, the condition that preserves the trace of the equation (\ref{secondForm}) is $\int_{0}^{t}\frac{\partial k(t',t)}{\partial t}dt'=0\implies\frac{\partial k(t',t)}{\partial t}=0$ for all $t'\in[0,t]$, which immediately indicates that the MKF is independent of $t$, \textit{i.e.,} $k(t',t)=k(t')$. Consequently, equation (\ref{secondForm}) reduces to 
\begin{equation}
    \dfrac{\partial\rho_{S}(t)}{\partial t} = \int_{0}^{t} dt' k(t') e^{\mathcal{L}t'}\circ\mathcal{E}\circ\mathcal{L}\rho_{S}(t-t').\label{finalForm}
\end{equation}
This is the final form of the \textit{post-Markovian master equation}. We now demonstrate that this time-nonlocal PMME interpolates between the Markovian ME and the exact Nakajima-Zwanzig equation. To establish this, we first consider $k(t')=\delta(t')$, implying that the additional measurement occurs only at the initial time $t'=0$ and no further measurements are performed during the intermediate times $(0,t)$. Under this condition, equation (\ref{finalForm}) simplifies to the Markovian GKSL ME, $\dfrac{\partial\rho_{S}(t)}{\partial t}=\mathcal{L}\rho_{S}(t)$. Again, by considering the form of the memory kernel superoperator as $\mathcal{K}(t')=k(t') e^{\mathcal{L}t'} \circ \mathcal{E} \circ \mathcal{L}$, equation (\ref{finalForm}) transforms into the time-nonlocal Nakajima-Zwanzig equation,
\begin{equation}
    \dfrac{\partial \rho_{S}(t)}{\partial t} = \int_{0}^{t}dt' \mathcal{K}(t')\rho_{S}(t-t').
\end{equation}
In the following section, we solve the PMME and present the condition for complete positivity.
\section{Solution of the master equation}
\label{solSection}
To solve the PMME (an integro-differential equation) (\ref{finalForm}), it is convenient to work in the eigenbasis of $\mathcal{L}$. In general, $\mathcal{L}$ is not a normal superoperator, \textit{i.e.,} $[\mathcal{L},\mathcal{L}^{\dagger}]\neq 0$,  which consequently leads to it possessing distinct sets of right and left eigenvectors, denoted by $\{R_{i}\}$ and $\{L_{i}\}$, respectively, which satisfy the eigenvalue equations $\mathcal{L}R_i=\lambda_iR_i$ and $L_i\mathcal{L}=\lambda_iL_i$. Note that both the sets are complete and mutually orthonormal with respect to the Hilbert-Schmidt inner product, \textit{i.e.,} ${\rm Tr}(L_iR_j)=\delta_{ij}$. Thus the state $\rho_{S}(t)$ can be expanded in the basis $\{R_{i}\}$ as 
\begin{equation}
     \rho_{S}(t)=\sum_{i=1}^{d^2}\mu_{i}(t)R_{i},\label{solExpansion}
\end{equation}
where $d = d_{S}$ is the dimension of the Hilbert space $\mathcal{H}_{S}$ of $S$, and the expansion coefficients satisfy $\mu_{j}(t)=\sum_{i}\mu_{i}(t){\rm Tr}(L_{j}R_{i})={\rm Tr}(L_{j}\rho_{S}(t))$.

\noindent Now, invoking equation (\ref{solExpansion}) in the PMME (\ref{finalForm}), we obtain
\begin{equation}
    \sum_{i=1}^{d^2} \dfrac{\partial\mu_{i}(t)}{\partial t} R_{i} = \sum_{i=1}^{d^2} \int_{0}^{t}dt' k(t') \mu_{i}(t-t') e^{\mathcal{L}t'}\circ\mathcal{E}\circ\mathcal{L}R_{i}.
\end{equation}
By multiplying both the sides by $L_{j}$ from the left and taking the trace, we get
\begin{align}
    \dfrac{\partial\mu_{j}(t)}{\partial t} &= \sum_{i=1}^{d^2} \lambda_{i} {\rm Tr}(L_{j}\circ\mathcal{E}\circ R_{i}) \int_{0}^{t} dt' k(t') e^{\lambda_{j}t'} \mu_{i}(t-t')\nonumber\\
    &= \sum_{i=1}^{d^2} \lambda_{i} {\rm Tr}(L_{j}\circ\mathcal{E}\circ R_{i})~k(t) e^{\lambda_{j}t}*\mu_{i}(t),\label{beforeLT}
\end{align}
where $*$ denotes a convolution. To obtain the first line, we utilized the relations $\mathcal{L}R_{i} = \lambda_{i}R_{i}$ and $L_{j} e^{\mathcal{L}t'}=e^{\lambda_{j}t'}L_{j}$.

\noindent Laplace transform (${\rm L_{T}}$) with respect to the parameter `$s$' transforms equation (\ref{beforeLT}) into
\begin{equation}
    s {\rm L_{T}} \left[\mu_{j}(t)\right]-\mu_{j}(0) = \sum_{i=1}^{d^2} \lambda_{i} {\rm Tr} \left(L_{j} \circ \mathcal{E} \circ R_{i}\right) {\rm L_{T}} \left[k(t) e^{\lambda_{j}t}\right] {\rm L_{T}} \left[\mu_{i}(t)\right].\label{afterLT}
\end{equation}
Equation (\ref{afterLT}) can be solved for the set $\left\{{\rm L_{T}}\left[\mu_{j}(t) \right] \right\}_{j=1}^{d^2}$ by formulating a matrix equation as follows. Substituting  $j=1,2,\cdots,d^2$ into equation (\ref{afterLT}), yields the following set of equations:
{\footnotesize
\begin{align}
    \left(s-\lambda_{1} {\rm Tr}\left(L_{1}\circ\mathcal{E}\circ R_{1}\right) {\rm L_{T}}\left[k(t)e^{\lambda_{1}t}\right]\right) {\rm L_{T}}\left[\mu_{1}(t)\right] - &\sum_{i\neq 1}^{d^{2}} \lambda_{i} {\rm Tr} \left(L_{1}\circ\mathcal{E}\circ R_{i}\right) {\rm L_{T}}\left[k(t) e^{\lambda_{1}t}\right]{\rm L_{T}}\left[\mu_{i}(t)\right] = \mu_{1}(0),\nonumber\\
     \left(s-\lambda_{2} {\rm Tr}\left(L_{2}\circ\mathcal{E}\circ R_{2}\right) {\rm L_{T}}\left[k(t)e^{\lambda_{2}t}\right]\right) {\rm L_{T}}\left[\mu_{2}(t)\right] - &\sum_{i\neq 2}^{d^{2}} \lambda_{i} {\rm Tr} \left(L_{2}\circ\mathcal{E}\circ R_{i}\right) {\rm L_{T}}\left[k(t) e^{\lambda_{2}t}\right]{\rm L_{T}}\left[\mu_{i}(t)\right] = \mu_{2}(0),\nonumber\\
     &\vdots\nonumber\\
     \left(s-\lambda_{d^{2}} {\rm Tr}\left(L_{d^{2}}\circ\mathcal{E}\circ R_{d^{2}}\right) {\rm L_{T}}\left[k(t)e^{\lambda_{d^{2}}t}\right]\right) {\rm L_{T}}\left[\mu_{d^{2}}(t)\right] - &\sum_{i=1}^{d^{2}-1} \lambda_{i} {\rm Tr} \left(L_{d^{2}}\circ\mathcal{E}\circ R_{i}\right) {\rm L_{T}}\left[k(t) e^{\lambda_{d^{2}}t}\right]{\rm L_{T}}\left[\mu_{i}(t)\right] = \mu_{d^{2}}(0),\label{solSet}
\end{align}
}
with $\mu_{j}(0) = {\rm Tr} \left(L_{j}\rho_{S}(0)\right)$.

\noindent From this set of equations (\ref{solSet}),  the corresponding matrix equation can be constructed as
\begin{equation}
    \Omega\mathcal{X} = \mathcal{Y}_{0},\label{matrixEquation}
\end{equation}
where $\mathcal{X} = \left({\rm L_{T}} \left[\mu_{1}(t)\right], {\rm L_{T}}\left[\mu_{2}(t)\right], \cdots, {\rm L_{T}} \left[\mu_{d^2}(t)\right]\right)^{T}$, $\mathcal{Y}_{0} = \left(\mu_{1}(0), \mu_{2}(0), \cdots, \mu_{d^2}(0)\right)^{T}$, and
{\footnotesize
 \begin{equation*}\label{matrix}
\renewcommand\arraystretch{1.8}
   \Omega=\begin{pmatrix}
    s-\lambda_{1} {\rm Tr}\left(L_{1}\circ\mathcal{E}\circ R_{1}\right) {\rm L_{T}} \left[k(t)e^{\lambda_{1}t}\right] & -\lambda_{2} {\rm Tr}\left(L_{1}\circ\mathcal{E}\circ R_{2}\right) {\rm L_{T}} \left[k(t)e^{\lambda_{1}t}\right]&\cdots&-\lambda_{d^2} {\rm Tr} \left(L_{1}\circ\mathcal{E}\circ R_{d^2}\right) {\rm L_{T}}\left[k(t)e^{\lambda_{1}t}\right]\\
-\lambda_{1} {\rm Tr} \left(L_{2}\circ\mathcal{E}\circ R_{1}\right) {\rm L_{T}} \left[k(t)e^{\lambda_{2}t}\right] & s-\lambda_{2} {\rm Tr} \left(L_{2}\circ\mathcal{E}\circ R_{2}\right) {\rm L_{T}} \left[k(t)e^{\lambda_{2}t}\right] &\cdots&-\lambda_{d^2} {\rm Tr}\left(L_{2}\circ\mathcal{E}\circ R_{d^2}\right) {\rm L_{T}} \left[k(t)e^{\lambda_{2}t}\right] \\
     \vdots& \vdots&\ddots&\vdots\\
-\lambda_{1} {\rm Tr} \left(L_{d^2}\circ\mathcal{E}\circ R_{1}\right) {\rm L_{T}} \left[k(t)e^{\lambda_{d^2}t}\right]&-\lambda_{2} {\rm Tr} \left(L_{d^2}\circ\mathcal{E}\circ R_{2}\right) {\rm L_{T}} \left[k(t)e^{\lambda_{d^2}t}\right]&\cdots &s-\lambda_{d^2} {\rm Tr}\left(L_{d^2}\circ\mathcal{E}\circ R_{d^2}\right) {\rm L_{T}} \left[k(t)e^{\lambda_{d^2}t}\right]\\
    \end{pmatrix}.
\end{equation*}
}
In general, $\Omega$ is assumed to be invertible, with its inverse denoted as $\Omega^{-1}$.

By solving the matrix equation (\ref{matrixEquation}), one can determine  $\left\{{\rm L_{T}}\left[\mu_{j}(t) \right] \right\}_{j=1}^{d^2}$. Applying the inverse Laplace transform, ${\rm L_{T}^{-1}}$, then yields ${\rm L_{T}^{-1}}\left\{{\rm L_{T}}\left[\mu_{j}(t)\right]\right\}_{j=1}^{d^2} = \left\{\mu_{j}(t)\right\}_{j=1}^{d^2}$, \textit{i.e.,}
\begin{equation}
    \mathcal{N} = {\rm L_{T}^{-1}} \left[\mathcal{X}\right] = {\rm L_{T}^{-1}} \left[\Omega^{-1}\right] \mathcal{Y}_{0} = \mathcal{W} \mathcal{Y}_{0},\label{solution_pre}
\end{equation}
here $\mathcal{N} = \left(\mu_{1}(t), \mu_{2}(t), \cdots, \mu_{d^2}(t)\right)^{T}$.

Finally, substituting $\mu_{i}(t) = \sum_{j=1}^{d^2}\mathcal{W}_{ij}(t)\mu_{j}(0)$ into equation (\ref{solExpansion}), where $\mathcal{W}_{ij}(t)$ denotes the time-dependent $ij$-th element of the matrix $\mathcal{W}$, yields the exact solution of the PMME (\ref{finalForm}) as
\begin{equation}
    \rho_{S}(t)=\sum_{i=1}^{d^2}\mu_{i}(t)R_{i} = \sum_{i=1}^{d^2}\sum_{j=1}^{d^2}\mathcal{W}_{ij}(t) \mu_{j}(0) R_{i}.\label{finalSolution}
\end{equation}
Therefore, obtaining the solution of the master equation (\ref{finalForm}) requires knowledge of the exact form of the MKF $k(t)$, the CPTP map $\mathcal{E}$ associated with the measurement, as well as the left and right eigenvectors along with their concomitant eigenvalues $\lambda_{i}$ of the Lindbladian $\mathcal{L}$.
\subsection{Dynamical map governing the PMME}
The solution presented in the equation (\ref{finalSolution}) can be expressed as
\begin{equation}
     \rho_{S}(t) = \sum_{i=1}^{d^2}\sum_{j=1}^{d^2} \mathcal{W}_{ij}(t) {\rm Tr} \left(L_{j}\rho_{S}(0)\right) R_{i} \equiv \Phi(t) \left[\rho_{S}(0)\right],
\end{equation}
where the dynamical map $\Phi(t)$ associated with the PMME is defined as 
\begin{equation}
    \Phi(t) \left[A\right] := \sum_{i=1}^{d^2}\sum_{j=1}^{d^2} \mathcal{W}_{ij}(t) {\rm Tr} \left(L_{j}A\right) R_{i}.\label{dynamicalMap}
\end{equation}
Assuming ${\rm det}(\mathcal{W})\neq 0$ for all $t$, we can impose that $\Phi(t)$ is invertible, with the inverse map defined as
\begin{equation}
    \Phi^{-1}(t) \left[B\right] := \sum_{i=1}^{d^2}\sum_{j=1}^{d^2} \mathcal{W}^{-1}_{ij}(t) {\rm Tr} \left(L_{j}B\right) R_{i},\label{Inverse_dynamicalMap}
\end{equation}
where, $\mathcal{W}^{-1}$ is the inverse of $\mathcal{W}$. The validity of the inverse dynamical map is established as follows
\begin{align}
    \Phi^{-1}(t) \circ \Phi(t) \left[X\right] &= \sum_{i,j=1}^{d^2} \mathcal{W}^{-1}_{ij}(t) {\rm Tr} \left(L_{j} \Phi(t) \left[X\right] \right) R_{i}\nonumber\\
    &= \sum_{i,j=1}^{d^2} \sum_{k,l=1}^{d^2}\mathcal{W}^{-1}_{ij}(t) \mathcal{W}_{kl}(t) {\rm Tr} \left(L_{l}X\right) {\rm Tr} \left(L_{j} R_{k}\right) R_{i}\nonumber\\
    &= \sum_{i,j,l=1}^{d^2} \mathcal{W}^{-1}_{ij}(t) \mathcal{W}_{jl}(t) {\rm Tr} \left(L_{l}X\right) R_{i} = \sum_{i=1}^{d^2} {\rm Tr} \left(L_{i}X\right) R_{i} = X.
\end{align}
In deriving the final expression, we utilized the property $\mathcal{W}^{-1}\mathcal{W}=\mathds{1}$, which implies that for each $i$ and $l$, $\sum_{j} \mathcal{W}^{-1}_{ij}(t)\mathcal{W}_{jl}(t) = \delta_{il}$.

Consequently, the time-nonlocal PMME (\ref{finalForm}) can be reformulated as a time-local differential equation in the time-convolutionless (TCL) form
\begin{equation}
    \dfrac{\partial\rho_{S}(t)}{\partial t} = \left[\int_{0}^{t} dt' k(t') e^{\mathcal{L}t'}\circ\mathcal{E}\circ\mathcal{L}\circ\Phi(t-t')\circ\Phi^{-1}(t)\right]\rho_{S}(t) \equiv \mathcal{K}_{TCL}(t)\rho_{S}(t),\label{TCL}
\end{equation}
where $\mathcal{K}_{TCL}(t)$ is a time-convolutionless generator.
\subsection{Complete positivity of the PMME}
Dependence on the MKF $k(t)$ may compromise the complete positivity of the post-Markovian dynamics. Therefore, it is crucial to determine the condition under which the PMME preserves complete positivity throughout the entire dynamical evolution. This analysis can be facilitated by employing the well-known Choi's theorem \cite{CHOI1975285}, which states that a map is completely positive if and only if its Choi matrix is positive semidefinite. The Choi matrix for a map $\Phi(t)$  is defined as $\sum_{k,l}\ket{k}\bra{l}\otimes\Phi(t)\left[\ket{k}\bra{l}\right]$. In the present context, the Choi matrix corresponding to the map defined in equation (\ref{dynamicalMap}) is given by
\begin{align}
    C_{\Phi(t)} &= \sum_{k,l}\ket{k}\bra{l}\otimes\sum_{i,j}\mathcal{W}_{ij}(t) {\rm Tr} \left(L_{j} \ket{k}\bra{l}\right) R_{i}\nonumber\\
      &= \sum_{i,j}\mathcal{W}_{ij}(t) \sum_{k,l} \bra{k} L_{j}^{\rm T}\ket{l} \ket{k}\bra{l}\otimes R_{i}\nonumber\\
      &= \sum_{i,j}\mathcal{W}_{ij}(t) L_{j}^{\rm T} \otimes R_{i}.
\end{align}
Therefore, the necessary and sufficient condition that the dynamical map (\ref{dynamicalMap}) associated with the PMME must satisfy to retain complete positivity throughout the entire evolution is
\begin{equation}
    C_{\Phi(t)} = \sum_{i,j}\mathcal{W}_{ij}(t) L_{j}^{\rm T} \otimes R_{i} \geq 0.\label{cptpCondition}
\end{equation}
To put it another way, equation (\ref{cptpCondition}) imposes an implicit constraint on the MKF $k(t)$ through $\mathcal{W}$, for a given Lindbladian $\mathcal{L}$.
\section{Thermalization}
\label{thermalSection}
With the rapid developments in collisional models, their significance extends beyond the theory of open quantum systems, finding remarkable applications in quantum thermodynamics \cite{PhysRevA.91.022121, PhysRevLett.115.120403, Pezzutto_2016, PhysRevX.7.021003, PhysRevA.98.032119, PhysRevLett.123.140601, Pezzutto_2019, PhysRevLett.123.180602, e21121182, PhysRevA.102.042217, PhysRevApplied.14.054005, Abah_2020, PhysRevResearch.2.033315, e22070763, Morrone_2023, PhysRevLett.130.200402, Saha_2024, ziman2}. In this section, we explore thermalization within the framework of CMs using the post-Markovian dynamics. In the context of post-Markovian dynamics, the post-measurement evolution is essentially Markovian, leading to the expected relaxation. Thermalization is a fundamental process in non-equilibrium thermodynamics. In simple terms, a system in contact with a thermal bath will ultimately relax toward thermal equilibrium. Specifically, let $\rho_{S}$ represent an arbitrary state of the system, $\eta_{B}$ the state of the thermal bath, and $\rho_{S}^{\star}$ the state of the system at thermal equilibrium. The process of thermalization is characterized by the following conditions: 

$(i)$ If the system is initially prepared in a state $\rho_{S}\neq\rho_{S}^{\star}$, the thermalization process ensures that, in the end, the composite state $\sigma_{SB}$ satisfies ${\rm Tr_{B}} \left(\sigma_{SB}\right) \simeq\rho_{S}^{\star}$ and ${\rm Tr_{S}} \left(\sigma_{SB}\right) \simeq\eta_{B}$, and 

$(ii)$ the state $\left(\rho_{S}^{\star}\otimes\eta_{B}\right)$ remains stationary under the thermalization dynamics.\\

In the context of Markovian CMs, thermalization has been studied through a more general process known as quantum homogenization\footnote{Quantum homogenization is a process in which the state of the system asymptotically approaches the identically prepared state of the ancillas.} \cite{ziman1, ziman2}. Recently, by modifying the Markovian model to include ancilla-ancilla interactions, Saha \textit{et al.} \cite{Saha_2024} have addressed the same problem within the framework of non-Markovian CMs. In both studies, the analysis was conducted specifically for the qubit case. Notably, the authors in \cite{Saha_2024} demonstrated that the thermalization rate of their non-Markovian dynamics could be matched to that of the Markovian scenario. More precisely, the thermalization time (\textit{i.e.,} the time required to reach a thermal state) for the non-Markovian dynamics is nearly identical to that of the Markovian counterpart but does not surpass it.\\

\begin{figure*}[ht!]
\begin{center}
    \includegraphics[width=0.70\textwidth,height=8cm]{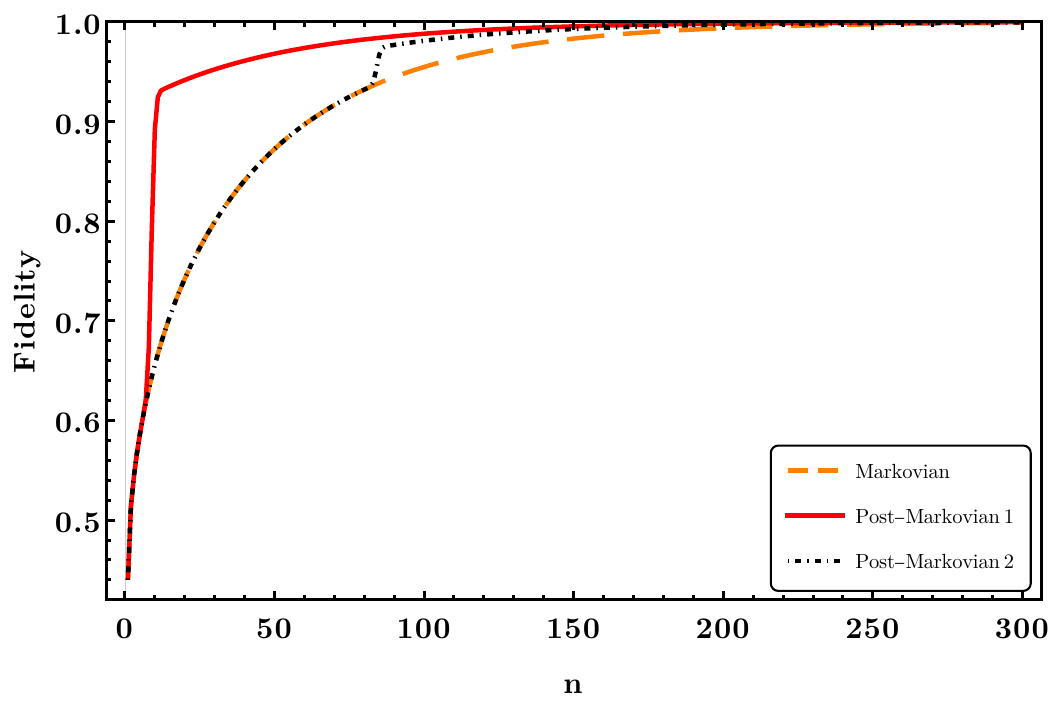}
    \caption{Plot of fidelity with the number of collisions $n$ illustrating thermalization for the post-Markovian dynamics and comparing it with the Markovian dynamics. Initial system state $\rho_{S}(0)$ (in pure state form) $=\frac{1}{\sqrt{5}}\ket{0}+\frac{2}{\sqrt{5}}\ket{1}$ and initial ancilla state $\eta = \frac{3}{5}\ket{0}\bra{0}+\frac{2}{5}\ket{1}\bra{1}$. In each case, $\alpha=0.1$, and $\beta=0.9$.}
    \label{thermalizationPlot}
\end{center}
\end{figure*}

Figure \ref{thermalizationPlot} demonstrates that, in the qubit case, thermalization is achievable under the post-Markovian dynamics. In figure \ref{thermalizationPlot}, we plot the fidelity between the reduced state of the system after each collision and the initial ancilla state $\eta$ with the number of collision $n$. The fidelity between two quantum states $\rho$ and $\sigma$ is defined as $F(\rho,\sigma) = \left\{{\rm Tr} \sqrt{\sqrt{\rho}\sigma\sqrt{\rho}}\right\}^{2}$ \cite{nielsen}. In figure \ref{thermalizationPlot}, the initial system state is chosen as the pure state $\rho_{S}(0) = \ket{\psi_{S}(0)}\bra{\psi_{S}(0)}$ with $\ket{\psi_{S}(0)} = \frac{1}{\sqrt{5}}\ket{0} + \frac{2}{\sqrt{5}}\ket{1}$, which exhibits coherence with respect to the computational basis $\left\{\ket{0}, \ket{1}\right\}$. Here, the computational basis is defined as the eigenstates of $\sigma_{z}$, where $\sigma_{z} \ket{0} = \ket{0}$ and $\sigma_{z} \ket{1} = -\ket{1}$. Whereas we consider the initial ancilla state to be $\eta = \frac{3}{5}\ket{0}\bra{0}+\frac{2}{5}\ket{1}\bra{1}$, which, with explicit reference to the ancilla Hamiltonian, can be interpreted as a thermal state with a fixed temperature. 
To generate the plot, we consider the collisional unitary to be the partial swap (PSWAP\footnote{In the qubit case, PSWAP is the unique system-ancilla unitary capable of achieving thermalization through homogenization \cite{ziman1}.}), given by $U(\alpha) = \cos\left({\alpha}\right) \mathds{1}_{4\times4} + i\sin\left({\alpha}\right) S_{4\times4}$. Similarly, the measurement unitary is chosen as $\mathcal{U_{M}} = U(\beta) = \cos\left({\beta}\right) \mathds{1}_{4\times4} + i\sin\left({\beta}\right) S_{4\times4}$, which is also a PSWAP. We assign weights to different measurements drawn from a Gaussian probability distribution. Two distinct scenarios are investigated for the post-Markovian dynamics: in the first, the measurement weights are predominantly confined to the early stages of the dynamics (red solid line), whereas in the second, they are primarily concentrated in the intermediate stages of the dynamics (black dot-dashed line). The plot clearly shows that the system, evolving under post-Markovian dynamics, indeed attains thermalization. In both of the above-mentioned scenarios, the system not only reaches the thermal state but does so at a faster rate compared to the Markovian counterpart (orange dashed line). Although this investigation employs non-selective measurements on the ancillas in the eigenbasis of $\sigma_{x}$, the thermalization profiles remain unchanged for different choices of measurement bases. This is because the measurement does not directly influence the dynamics; instead, its probabilistic nature induces bath memory effects, thereby modifying the otherwise Markovian dynamics. Between the two scenarios for post-Markovian dynamics, we notice that the former, in which maximum weights of measurements are confined to the early stages of the dynamics, results in faster thermalization (although marginally) than the latter, where they are concentrated in the intermediate stages. For any initial states of the system and ancilla, all the aforementioned observations hold true for the specified collisional and measurement unitaries, although the thermalization rate may vary. To ensure a fair comparison across all cases, the values of the parameters \textit{i.e.,} $\alpha$ and $\beta$ are kept unchanged.
\section{Conclusion}
\label{conclusion}
In this paper, we have derived a post-Markovian master equation starting from a microscopic collisional model framework. To incorporate bath memory into an otherwise Markovian collisional model, we have employed probabilistic measurements on the ancillas in between the dynamics. These memory effects are encapsulated by the phenomenologically introduced memory kernel function in the resulting master equation. Although the initial states of the ancillas are assumed to be pure in deriving the master equation, the form of the equation remains unchanged when the ancillas are initialized in mixed states. This robustness arises from the fact that any mixed state can be expressed as a convex combination of pure states. The resulting PMME interpolates between two extreme situations: a purely Markovian ME and the exact Nakajima-Zwanzig equation. An analytical solution of the PMME is presented. The presence of the MKF hinders the unconditional complete positivity of the dynamics. We have specified a necessary and sufficient condition on the MKF to ensure the retention of complete positivity throughout the entire dynamical evolution. The derivation of the PMME through CM highlights the versatility of these models in representing a broad range of physical open quantum system dynamics. Consequently, this collisional model-based approach to post-Markovian dynamics provides a basis for simulating certain classes of non-Markovian dynamics on suitable physical platforms. Subsequently, qubit thermalization under the post-Markovian dynamics is also examined. Surprisingly, post-Markovian dynamics is found to accelerate the thermalization process, achieving rates faster than those of the Markovian counterpart. Consequently, implementing this thermalization protocol under post-Markovian dynamics in the thermalization strokes of various heat engines would enhance their performance.

\section*{Acknowledgements}
TS would like to thank Dr. Arpan Das for many fruitful discussions and suggestions.

\bibliographystyle{JHEP}
\bibliography{postMarkovCM}

\end{document}